\documentclass{sigcomm-alternate}
\usepackage{amsmath,amstext,verbatim,epsfig}
\usepackage{colordvi,pstcol}
\usepackage{graphicx,psboxit}
\usepackage{psfrag}
\newcommand\GG{\mathcal{G}}

\newtheorem{remark}{Remark}

\begin{document}
\title{Statistical reliability and path diversity based PageRank algorithm improvements}

\numberofauthors{1}
\author{
  \alignauthor Dohy Hong\vspace{2mm}\\
  \affaddr{Alcatel-Lucent Bell Labs}\\
  \affaddr{Route de Villejust}\\
  \affaddr{91620 Nozay, France}\\
  \email{\normalsize dohy.hong@alcatel-lucent.com}
}

\date{\today}
\maketitle

\begin{abstract}
In this paper we present new improvement ideas of the original PageRank algorithm.
The first idea is to introduce an evaluation of the statistical reliability of the ranking score
of each node based on the local graph property and the second one is to 
introduce the notion of the path diversity.
The path diversity can be exploited to dynamically modify the increment value of 
each node in the random surfer model or to dynamically adapt the damping factor.
We illustrate the impact of such modifications through examples and simple simulations.
\end{abstract}
\begin{psfrags}
\section{Introduction}\label{sec:intro}
There was an important research investment during at least 10 years
on PageRank algorithm and related topics 
(cf. \cite{rad, eigenfactor, hotho, poprank, deeper, video, balmin, xrank, avra2, bian, yen}), 
but there were few results concerning the original PageRank algorithm modification.

PageRank is a nice solution to evaluate the importance of the nodes
of a graph based on the resolution of a fixed point problem
associated to the random surfer model and to the Markov chain associated 
to the random walk.
The PageRank algorithm can be then seen as a Perron-Frobenius problem
(simplified formulation):
$$
A.X = X
$$
where $A$ is the transition matrix associated to the random surfer model
(the size of the state is $N$, if there are $N$ URLs)
$$
A(i,j) = \frac{\epsilon}{N} + (1-\epsilon) \sum_{j=1}^N \frac{1_{j\to i}}{N(j)}
$$

and 
$X$ the stationary probability. $X$ measures the relevancy of each URL
(cf. \cite{page, hits, deeper}), which is proportional
to the average sojourn time at each node during the random walk.

In this paper, we are interested in investigating one very specific
issue which may be the Achilles' heel of PageRank.
This issue is related to the possible impacts from the choice of the
damping factor \cite{damp, damp2, avra, bressan, boldi}.
The role of the damping factor in the initial PageRank algorithm
can be associated in the random surfer model to the probability that
the surfer gets bored after several clicks and switches to a random page.
More technically speaking, it may have three roles:
\begin{itemize}
\item{[Irreducibility]} firstly, it plays a role of mixing all nodes and making the associated
  Markov chain irreducible (i.e. we have a single connected component);
\item{[Indirect inheritance]} secondly, it controls directly the way the importance weights are inherited when
  following the links (cf. illustration in Figure \ref{fig:graphInherit});
  as a consequence, it impacts the global ranking results (cf. \cite{damp,damp2});
\item{[Trap nodes]} finally, it avoids the random walk staying too long in a trap position; 
  the trap position can be one node (loop) or a group of nodes from which
  the outbound links are all local; the damping factor would enable to leave such a
  position and explore the whole space.
\end{itemize}

Because of the second point above, we think that the damping factor could partially
induce an arbitrary ranking results, which may be undesirable.
This is further illustrated in Figure \ref{fig:graphInherit}:
the node $i_2$ inherits most of scores from $i_1$. 
More precisely, if the score of $i_1$ is $C(i_1)$, $i_2$
inherits from $i_1$: $C(i_1)/2\times (1-\epsilon)$, where $1-\epsilon$ is the damping factor
and $C(i_1)$ is divided by the number of outgoing links from $i_1$.
Therefore, when applying PageRank family approaches
this indirect influence depends directly on the value of the damping factor.

\begin{figure}[htbp]
\centering
\psfrag{i1}{$i_1$}
\psfrag{i2}{$i_2$}
\includegraphics[width=7cm]{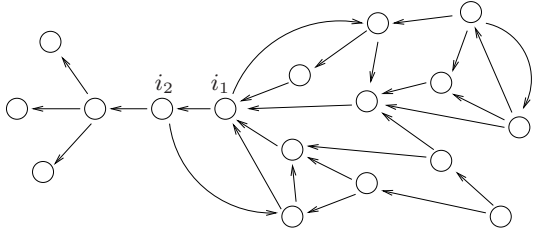}
\caption{$i_2$ inherits scores from $i_1$}
\label{fig:graphInherit}
\end{figure}

In the next sections, we present how we can correct or at least control such an impact
based on the idea of the statistical reliability (Section \ref{sec:qualif}) and on the idea
of the path diversity (Section \ref{sec:diversity}).

\section{Statistical reliability}\label{sec:qualif}
We consider the random walk model on a graph $\GG$ of $N$ nodes
where each transition from node $j$ to node $i$ is defined by $p(i,j)$.
In particular, we focus on a homogeneous graph (one could extend
the same approach on a heterogeneous graph) where $p(i,j)$ is
defined by $N(j)$ the number of outgoing links from node $j$
(if there is a link from $j$ to $i$):
$$
p(i,j) = \frac{\epsilon}{N} + \frac{1-\epsilon}{N(j)}
$$
where $1-\epsilon$ is the commonly called damping factor \cite{page}.

In this section, we assume we already solved the original PageRank equation
\begin{eqnarray}\label{eq:PR}
X(i) &=& \frac{\epsilon}{N} + (1-\epsilon) \sum_{j=1}^N \frac{1_{j\to i}X(j)}{N(j)}
\end{eqnarray}
to find the ranking of each node $i$ of the graph , $X(i)$ defining
the importance weight (real value between 0 and 1) of the node $i$.

Now, we introduce a method to evaluate the statistical reliability of $X(i)$
based on the distribution of the local incoming links's contribution.

\subsection{General expression}
We assume that the Markov chain associated to the random walk
is described by the transition probability $P$ with $p(i,j) = P_{i,j}$
(probability to jump from $j$ to $i$)
and its stationary probability $X = (x_1,...,x_N)$.

Then we define the quantity $r(i,j)$ by:
\begin{eqnarray}\label{eq:rrr}
r(i,j) &=& p(i,j)\times x_j/x_i
\end{eqnarray}

By definition $\sum_{j=1}^N r(i,j) = 1$ and $r(i,j)$ can be simply
interpreted as the contribution of $j$ on $x_i$.

We define the following quantity measuring say the statistical error on $x_i$:
\begin{eqnarray}
E(i) &=& \sum_{j=1}^{N}\left( r(i,j) \right)^\alpha \times \beta
\end{eqnarray}
with $\alpha > 1$ and $\beta\in [0,1]$ (for instance, $\beta = 0.5$ or $1$ and $\alpha = 2$
which seems to be the most natural choices).
And we define:
\begin{eqnarray}
F(i) &=& 1- E(i).
\end{eqnarray}

The function $F(i)$ can be interpreted as what we called
the statistical reliability measure of $x_i$:
$F(i)$ is close to one when the $x_i$ is obtained from an equal contribution
of a large number of incoming links ($F(i) = 1-\beta/n$, if equal contribution from $n$
links),
whereas when the distribution of $r(i,j)$ is concentrated
on a single node $j$, $F(i)$ becomes close to $1-\beta$ which is its minimum value.

\begin{remark}
In the computation of $F$, we can include or not the transition probability
resulting from the damping factor; however, it seems more natural to
exclude it, since this is an artefact introduced for the computation
and is not part of human built links.
\end{remark}
\begin{remark}
The function $F(i)$ can be also interpreted as an evaluation of the robustness
of the score $x_i$, if for instance one local incoming link should be dropped.
\end{remark}

\subsection{Random walk based counters}\label{sec:rw}
Here, we assume that we maintain a counter vector $C$ of size $N$ to count the number
of visits of all nodes during the random walk. 
We define a counter matrix $R$ of size $N\times N$ and
we increment the counter $R(i,j)$ by one when we jump from $j$ to $i$ node.
If we call $C(i)$ the counter associated to the node $i$, the ratio
$r(i,j) = R(i,j)/C(i)$ gives the contribution ratio of node $j$ on $C(i)$.
Then, we can define $F(i)$ function as above.

Based on the formula \eqref{eq:rrr}, one may adapt the computation of $r(i,j)$
when other strategies are used to solve the PageRank equation \eqref{eq:PR}.

\subsection{Exploitation of the reliability function $F$}
The function $F$ can be exploited for several purposes:
\begin{itemize}
\item for the visualization issue: when we need to show the $K$ most
  relevant nodes associated to a node $i$, we can select the $K$ nodes
  based on the $K$ highest ratio $r(i,j)$; this idea can be generalized
  taking into account all distant neighbour nodes by summing
  the products of the form
  $r(i,j_1)\times r(j_1,j_2)...\times r(j_n, j)$ to a distant node $j$
  considering all possible path to $i$; in such a generalization, we could also
  take into account the damping factor by multiplying by $(1-\epsilon)^n$ depending
  on the length of the path $n$;
  for the nodes pointed
  directly or indirectly by $i$, we can obviously use $x_i$;
\item the distribution of $r(i,j)$ may be interpreted as a statistical signature of
  the ranking of $i$ and can be used to qualify the node $i$'s ranking
  and even more it can be used to modify the ranking value itself (cf. Section \ref{sec:rank}).
\end{itemize}

\section{Path diversity}\label{sec:diversity}
Here we define the notion of the path diversity to differentiate the increment
value (for the random walk based counters $C$, see Section \ref{sec:rw}). 
The main motivation of this approach is to avoid to give too high importance
to terminal or trap positions without being forced to play with the damping
factor which may have other global effect (such as what we called indirect inheritance
in Section \ref{sec:intro}).

We could use more or less {\em aggressive} definition of the path diversity.
Here we give three different formulations.

\subsection{Path diversity PD1}
This is the mildest version:
we keep a memory of the $L$ (if the length of the path from the initial
position or the last reinitialized node is less than $L$, we take the path length from
this position for $L$)
last visited nodes $LP = (n_1,...,n_L)$ (where $n_1$ is the last recently
visited node) and define the path diversity $div(LP)$ by the equation:
\begin{eqnarray}\label{eq:diversity1}
div(LP) &=& \frac{\sum_{i=1}^L f(i)\times g(i)}{\sum_{i=1}^L f(i)},
\end{eqnarray}
where $f$ and $g$ can be defined in two ways:\\

Power-law model:
\begin{eqnarray}\label{eq:lm}
f(i) &=& 1/i^\alpha,\\
g(i) &=& 1 - \sum_{j>i, n_i=n_j}\frac{1}{(j-i)^\alpha}.
\end{eqnarray}

Exponential model:
\begin{eqnarray}\label{eq:em}
f(i) &=& \gamma^i,\\
g(i) &=& 1 - \sum_{j>i, n_i=n_j}\delta^{j-i}
\end{eqnarray}
where $\delta$ should be less than 0.5. The specific choice of $\delta = 0.5$
seems the most interesting candidate (if $L$ is very large and the path
is a local loop on a same node, $g(i)$ would tend to zero).

\subsection{Path diversity PD2}
Here, we assume that
we keep memory of the full path from the last reinitialization time (due to terminal
positions or application of damping factor).
If the last visited nodes are $LP = (n_1,...,n_L)$ (where $n_1$ is the last recently
visited node) and the current position is $n_0$, we define the path redundancy of depth $i$ as:
\begin{eqnarray*}\label{eq:diversity2}
red(i) &=& g(i), \mbox{ if LP includes a path of length $i$ with a} \\
&& \mbox{ first node equal to $n_0$ and the last node}\\
&& \mbox{ equal to $n_i$,}\\
 & =& 0, \mbox{ otherwise}
\end{eqnarray*}
and the path diversity as:
\begin{eqnarray}\label{eq:pdiversity2}
div(LP) &=& \max(\sum_{1}^{\infty} g(i) - \sum_{1}^{L} red(i),0).
\end{eqnarray}

Function $g$ can be defined in different ways, in particular we
can define two types of model:\\

Power-law model:
\begin{eqnarray}\label{eq:lm2}
g(i) &=& \frac{1}{i^\alpha}.
\end{eqnarray}

Exponential model:
\begin{eqnarray}\label{eq:em2}
g(i) &=& \frac{1}{\gamma^i}.
\end{eqnarray}
The specific choice of $\gamma = 2$
seems to be an interesting natural candidate (so that the increment is equal to 1
for a {\em full} diversity, i.e. all nodes are different).

\subsection{Path diversity PD3}
This one is probably the most {\em aggressive} version of diversity:
this is as PD2, but with the following formulation:

\begin{eqnarray*}\label{eq:diversity3}
div(LP) &=& 0, \mbox{ if there is a node $n_i$ equal to $n_0$,}\\
 & =& 1, \mbox{ otherwise.}
\end{eqnarray*}

However, the impact of PD3 if we continue the random walk is not
clear (we may have $div(LP)$ equal to zero than equal to 1). And such
a definition would be more relevant associated with ideas of
Section \ref{sec:prd}.

\begin{remark}
The common intuition of formulas above is to define a function that decreases as
the number of unique elements in $LP$ is small
and with a higher impact when the
duplicated node position is closer to the current position (PD1)
or when the size of the duplicated jump is small (PD2).
\end{remark}

\begin{remark}
The notion of path diversity is natural in the context of the random
walk. If the PageRank equation is to be solved/computed differently, an
adaptation of this approach may not be feasible and/or introduce an
additional computation cost.
\end{remark}

\begin{remark}
In a practical solution, PD1 should have a minor impact on the global
ranking, whereas PD3 will penalize the most the trap positions.
With the usually applied value of the damping factor (i.e. 0.85), the
depth of the graph traversal before reinitialization is small, hence PD3
definition can make sense in most of situations in a large graph.
\end{remark}

\subsection{Other application of the path diversity}\label{sec:prd}
Another possible way to exploit the path diversity is to take the
damping factor as a function of the path diversity, for instance
with PD1, PD2 or PD3.

A simple concrete solution can be: set $\epsilon = 0$ if the current position
has been already visited in the past (from the last reinitialization time): 
in Section \ref{sec:simu}, we show some
results of such a strategy we called PR$+$D.

\section{Ranking modification}\label{sec:rank}
\subsection{Reliability based modification}
We propose a new adaptation of PageRank
replacing the initial ranking $X(i)$ as follow (PR$\times$F):

\begin{eqnarray}\label{eq:reliability}
X'(i) &=& F(i) \times X(i).
\end{eqnarray}

One simple motivation of such a modification is to differentiate
the case when the $i$'s ranking is mostly inherited from a very small number
of significant neighbour nodes (say Dirac type distribution)
from the case when the contribution from the neighbour
nodes are spread on a large number of them (more uniform distribution).
From such an information, one may decide (depends of course on the context)
to credit more scores (or importances) on the nodes that depends more uniformly on a large
number of nodes, which could be also a sign of the consensus on the ranking.


With such a modification, it makes sense to use a smaller damping factor
than 0.85.

We further illustrate this in two simple examples below.

\subsubsection{Example case C1}
We set $\alpha = 2$, $\beta = 0.5$ and $\epsilon = 0.15$.
In Figure \ref{fig:graphInhc1}, if we assume $a$ has a reference score of 4
and $c$ a score of 3 (up to a constant multiplicative factor),
$b$ inherits from $a$: $4\times 0.85 = 3.4$ which is higher than $c$.

\begin{figure}[htbp]
\centering
\includegraphics[width=5cm]{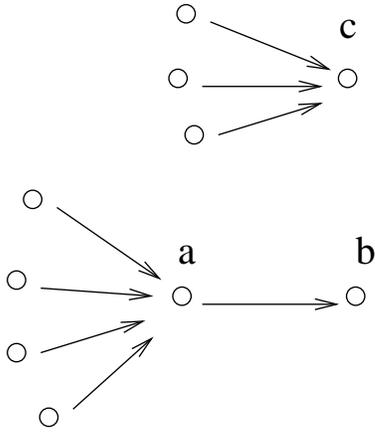}
\caption{$b$ inherits scores from $a$}
\label{fig:graphInhc1}
\end{figure}

Applying the reliability function $F$: we get for $a$: $4\times (1 -\beta/4) = 3.5$,
$b$: $3.4\times(1-\beta) = 1.8$
and for $c$: $3\times (1 -\beta/3) = 2.5$.

\subsubsection{Example case C2}
We set $\alpha = 2$, $\beta = 1$ and $\epsilon = 0.15$.
In Figure \ref{fig:graphInhc2}:
From initial PageRank: $b$ inherits from $a$ (assuming a score 6 for $a$): $6\times 0.85 = 5.1$.
Then within $b, c, d$, the average sojourn time before the reinitialization is
$1/0.15$ to be shared between the 3 nodes. Therefore, we have for
$(a,b,c,d)$: $(6, 7.3, 2.2, 2.2)$.
\begin{figure}[htbp]
\centering
\includegraphics[width=7cm]{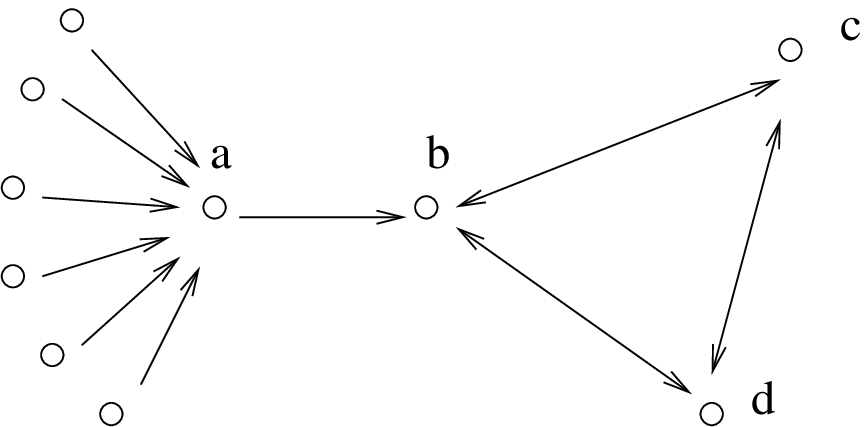}
\caption{Case C2}
\label{fig:graphInhc2}
\end{figure}

Applying the reliability function $F$: we get for $(a,b,c,d)$: 
$(5, 3.4, 1.1, 1.1)$.\\

Now with $\epsilon = 0.05$:
with the initial PageRank we obtain: $(6, 12.4, 6.7, 6.7)$ and 
applying the reliability function we have: $(5, 8.0, 3.3, 3.3)$.
This scenario shows also the necessity of keeping the damping factor
not too small to avoid the deadlock position in $b, c, d$ which tends
to overestimate their importance.\\

The modification proposed in the next section should allow one
to take $\epsilon$ close to zero without putting a too big importance weight on the nodes $b, c, d$.

\subsection{Path diversity based modification}
Here we illustrate the impact of the introduction of the path diversity
in the original PageRank equation.

\subsubsection{Example case C1}
In this simple case, all visited nodes (before a reinitialization is
required when reaching nodes $b$ or $c$) are different. Therefore,
the path diversity $div(LP)$ should be constant and does not impact
the ranking.

\subsubsection{Example case C2}
In this scenario, when $\epsilon$ is close to 0, $b, c, d$ become a trap position
and their importance weights should asymptotically sum up to 1.
With PD1, the importance weight of $a$ tends to zero as well (not aggressive enough).
Introducing the path diversity with PR2 or PD3, 
even if $\epsilon$ is equal to 0, the increment values
for $b, c, d$ quickly tend to zero, guaranteeing a strictly positive
weight of $a$ (and of other nodes). 
With P$+$D, the importance weights of $b, c, d$ are the most penalized.

\section{Simulation results}\label{sec:simu}
Here we set a simple simulation scenario to get a first evaluation of our
proposed solution and comparison to the original PageRank approach on the web graph.
We don't pretend to generate any realistic model, for more details on the
web graph the readers may refer to \cite{broder, kumar, barabasi, barabasi00, levene}.

\subsection{Scenario}
We set $N$ the total number of nodes (URLs) to be simulated.
Then we create $L$ random links (directional) to connect a node $i$
to $j$ as follow:

\begin{itemize}
\item the choice of the source node is done following a uniform sampling in Scenario S1
  or following a power-law: $1/k^\alpha$ in Scenario S2;
\item the choice of the destination node is done following a power-law: $1/k^\alpha$.
\end{itemize}

\begin{figure}[htbp]
\centering
\includegraphics[width=6cm,angle=-90]{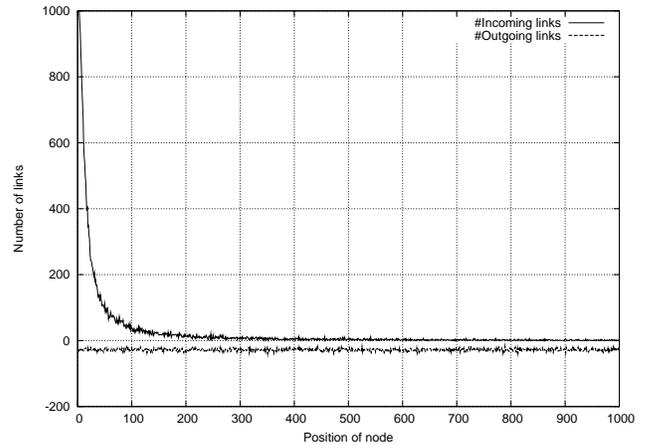}
\caption{Scenario S1: number of incoming and outgoing links: 27880 links created.}
\label{fig:S1}
\end{figure}

For simplicity, we assumed no correlation between the number of incoming and outgoing
links: in Scenario 1, a uniform sampling does not introduce correlation.
In Scenario 2, we first associate to node $k$ a probability proportional to $1/k^\alpha$
followed by a large number (by default $N$) of permutations of randomly chosen pair 
of nodes $(i,j)$: the final results define the {\em randomized} probability to be chosen
as a source node.
In both scenarios, we order the $N$ nodes by its popularity (probability to
be chosen as destination node), associating a probability proportional to $1/k^\alpha$ to
the node position $k$: in the following, we call this the native order which is
very close to the ordering by the number of incoming links (and not equal because of the
random realization).

\begin{figure}[htbp]
\centering
\includegraphics[width=6cm,angle=-90]{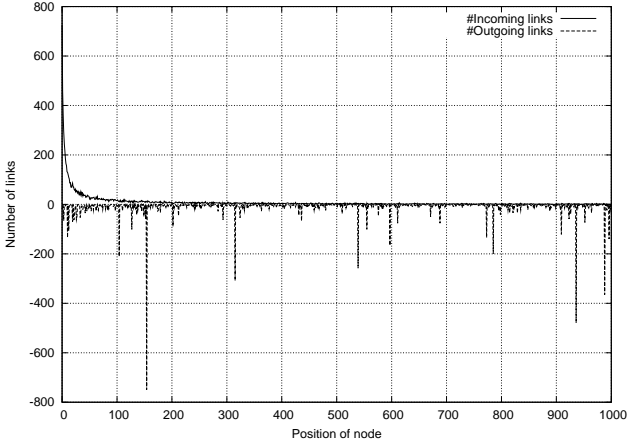}
\caption{Scenario S2: number of incoming (same as S1) and outgoing links ($N$ random permutations): 9533 links created.}
\label{fig:S2}
\end{figure}

When the link already exists between the source and destination nodes, we don't
modified anything (that's why we have less than $L$ links created).
The consequent results on the number of incoming and outgoing links are shown in
Figures \ref{fig:S1} and \ref{fig:S2} ($N=1000$ and $L=100\times N$, $\alpha=1.5$).

Figure \ref{fig:S1b} shows the power-law on the number of incoming links (by construction) and
on the number of nodes with $k$ incoming links (as a consequence).

\begin{figure}[htbp]
\centering
\includegraphics[width=6cm,angle=-90]{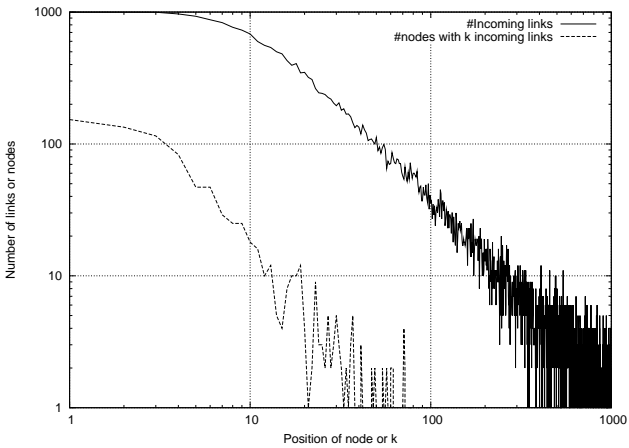}
\caption{Scenario S1/S2: number of incoming links (w.r.t. node position) and number of nodes with k incoming links 
(w.r.t. k) in logscale. $\alpha=1.5$.}
\label{fig:S1b}
\end{figure}

\subsection{Analysis}
\subsubsection{Scenario S1}
Figure \ref{fig:S1-r1} shows the results of PageRank based ranking relevancy scores
of the $N$ nodes: in this case, the PageRank ranking follows closely the number of
incoming links based ordering and the application of the function $F$ merely modifies
the results (we can only notice a bit more smoothed curve): 
because the choice of the source nodes is made randomly, the differentiation
of the $N$ nodes are only based on their difference on the probability to be chosen as
destination node. So we can consider here the ranking based on the number of
incoming links as the theoretically optimal one.

\begin{figure}[htbp]
\centering
\includegraphics[width=6cm,angle=-90]{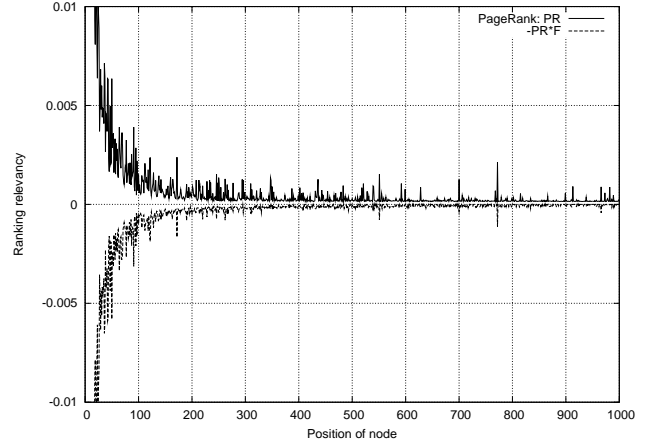}
\caption{Scenario S1: PR and PR$\times$F.}
\label{fig:S1-r1}
\end{figure}

To evaluate the difference of the ranking scores of two ranking approaches R1 and R2
(associated to their respective normalized relevancy scores $X_1()$ and $X_2()$),
we define the average deviation by:
$$
dev(R1,R2) = \frac{1}{N}\sum_{i=1}^N |\sum_{k=1}^{i} X_1(k)-X_2(k)|
$$
where $\sum_{k=1}^{i} X_1(k)$ gives the importance score of the $i$ first
nodes (following the native order).

\begin{figure}[htbp]
\centering
\includegraphics[width=6cm,angle=-90]{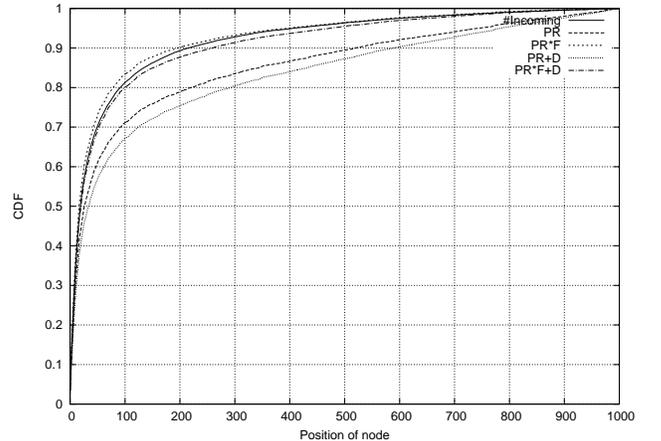}
\caption{Scenario S1: CDF of \#incoming, PR, PR$\times$F, PR$+$D and PR$\times$F$+$D.}
\label{fig:S1-dev-bis}
\end{figure}

In Table \ref{tab:onebis}, we show the average deviation (w.r.t. the number of incoming links)
for the four approaches. The application of $F$ makes the scores much more closer
to the ordering scores based on the number of incoming links,
which is expected since the factor $F$ will tend to favour the nodes receiving more incoming links.
The application of the diversity (PR$+$D, cf. Section \ref{sec:prd}) mainly reduces here the score of the best ranked nodes and
this explains its higher average deviation compared to PR. 

\begin{table}[hbtp]
\begin{center}
\begin{tabular}{|l||c|c|c|c|}
\hline
$\alpha$  & PR & PR$\times$F & PR$+$D & (PR$+$D)$\times$F\\
\hline 
1.5 & 0.062 & 0.0055 & 0.084 & 0.0077\\
\hline
2.0 & 0.071 & 0.0082 & 0.12 & 0.0047\\
\hline
2.5 & 0.073 & 0.0028 & 0.14 & 0.0017\\
\hline
\end{tabular}\caption{Average deviation.}\label{tab:onebis}
\end{center}
\end{table}

To better highlight the difference of the ranking scores of two approaches,
a node level deviation measure is defined as follows: given two ranking approaches R1 and R2,
we first evaluate node per node its relevancy score ratio to the average value by:
$Y(i) = X(i)\times N/\sum_i X(i)$ for R1 and R2;
then we measure the deviation between R1 and R2
by $dev(i) = Y_2(i)/Y_1(i)$.
The results are shown in Figure \ref{fig:S1-dev}: we compared PR and PR$\times$F to the ranking score based on
the number of incoming links. 
We observe more clearly the fact that the deviation is much more reduced when the function
$F$ is applied: this node level deviation evaluation allows one to easily observe the differences at
different ranking scale.
We also see that the deviation is naturally higher when there are more noises
(when the number of incoming links decreases).
\begin{figure}[htbp]
\centering
\includegraphics[width=6cm,angle=-90]{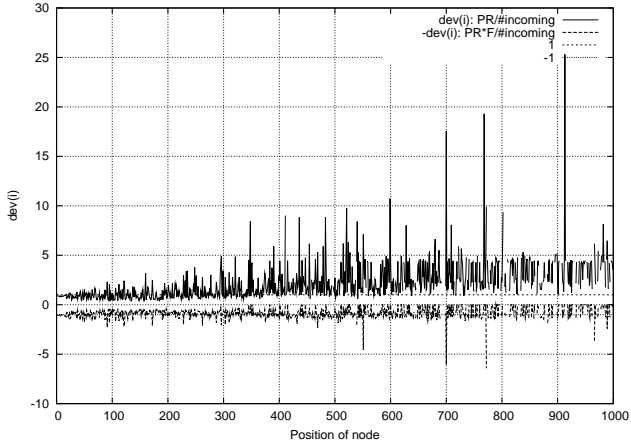}
\caption{Scenario S1: deviation measure. (PR/\#incoming) and (PR$\times$F/\#incoming).}
\label{fig:S1-dev}
\end{figure}


\subsubsection{Scenario S2}
Figure \ref{fig:S2-r1} shows the results of PageRank based ranking relevancy scores
of the $N$ nodes: in this case, there are 5 visible nodes after position 100 having a
very good relevancy score with PR. Because of the relevancy inheritance of PR,
even if they have few incoming links, their scores are very high when they are pointed
by the best ranked nodes distributing few outgoing links.
We can see that with the application of $F$ this effect disappears.

\begin{figure}[htbp]
\centering
\includegraphics[width=6cm,angle=-90]{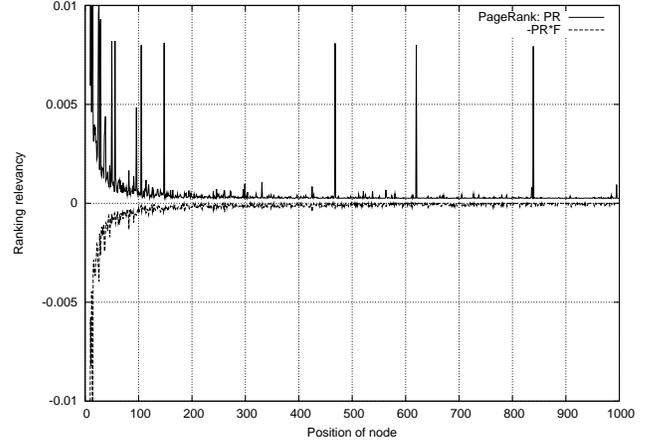}
\caption{Scenario S2: PR and PR$\times F$.}
\label{fig:S2-r1}
\end{figure}

The comparison of PR and PR$\times$F for the deviation measure is shown in Figure \ref{fig:S2-dev}.
We clearly see the big deviations with the 5 nodes we mentioned above.

\begin{figure}[htbp]
\centering
\includegraphics[width=6cm,angle=-90]{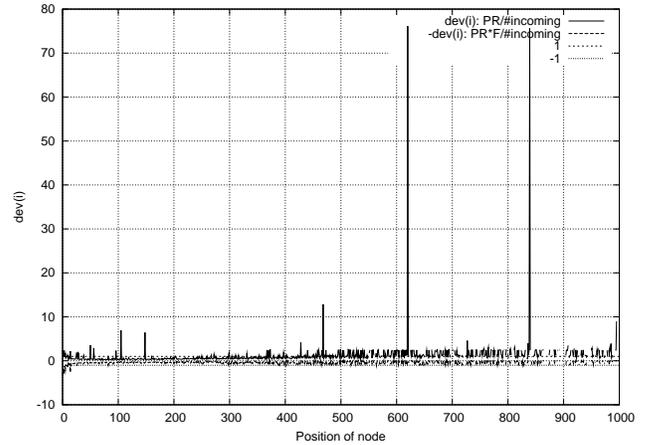}
\caption{Scenario S2: deviation measure. (PR/\#incoming) and (PR$\times$F/\#incoming).}
\label{fig:S2-dev}
\end{figure}

\subsubsection{Scenario S2b}
The scenario S2b is as S2, except for the node 1 we imposed one outgoing link to
the node 100 and the node 100 has also an unique outgoing link pointing to itself.
This scenario is meant to illustrate the impact of the trap position and how we
can control this impact.

Figure \ref{fig:S2b-r1} shows the results of PR and $PR\times F$. Both results
shows a very high relevancy score of the node 100.
In fact, the relevancy score of the node 100 (with PR) mainly comes
from the inheritance from the node 1 which can be estimated
by $PR(1)\times 0.85/0.15 = 0.08785\times 0.85/0.15 = 0.4978$
(which is very close to $x_{100} = 0.4993$). Node 100 has 17 incoming links,
but the main contribution is from the node 1 and as a consequence
it has a small reliability score of 0.25. This reduced score decreased
the importance of the node 100 (from $0.5$ to $0.13$), but can not
control the effect of the self-pointing influence.

\begin{figure}[htbp]
\centering
\includegraphics[width=6cm,angle=-90]{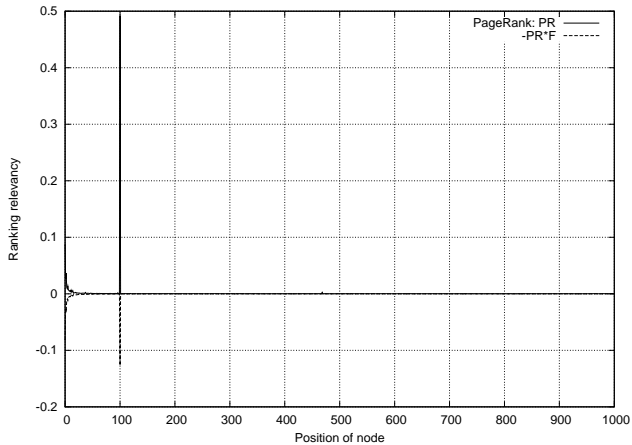}
\caption{Scenario S2b: PR and PR$\times$F.}
\label{fig:S2b-r1}
\end{figure}

In Figure \ref{fig:S2b-r2}, we show the results of PR$+$D:
here we see that with PR$+$D, we suppressed the self-pointing influence, but still
the node 100 inherits from the first node a score of: $0.15*0.85 = 0.13$ (the score
of the node 1 is of course modified by modifying the damping factor value dynamically).

Now applying the function $F$ on PR$+$D, we see that the node 100 is no more
differentiated.

\begin{figure}[htbp]
\centering
\includegraphics[width=6cm,angle=-90]{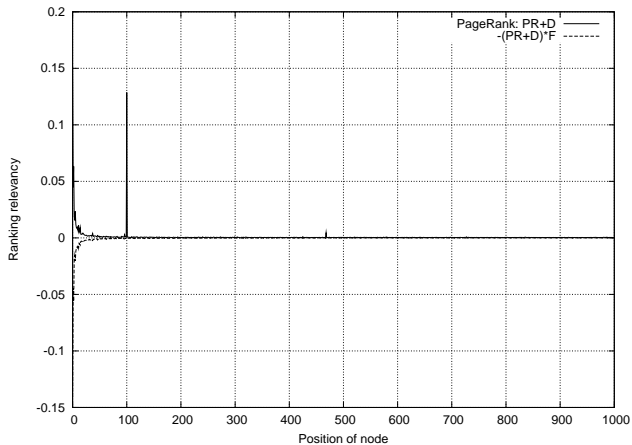}
\caption{Scenario S2b: PR$+$D and (PR$+$D)$\times$F.}
\label{fig:S2b-r2}
\end{figure}

Here we showed a rather extreme case, with a maximum penalty
with $F$ ($\beta = 1$) and with a maximum penalty from a non-diversity
to show how much it can impact and more importantly to illustrate the
fact that we can keep a control on what we called {\em indirect inheritance}
and the impact of {\em trap nodes} thanks to our modifications.
In a practical solution, it is necessary to correctly tune these values.

\section{Conclusion}\label{sec:conclusion}
In this paper, we defined the statistical reliability function associated to each
node of the graph and showed how it can be applied to possibly improve
the initial algorithm of PageRank results.
We also discussed the benefit of introducing the notion of the path diversity
to modify the increment value during the random walk or to modify the damping
factor. 
We showed the possible consequences through simple simulation scenarios.

In a future work, we expect to test/validate those ideas through a real data
based evaluation.

\end{psfrags}
\bibliographystyle{abbrv}
\bibliography{sigproc}

\end{document}